\documentstyle[12pt]{article}
\newlength{\extraspace}
\newlength{\extraspaces}
\newcommand{\be}{\begin{equation}
\addtolength{\abovedisplayskip}{\extraspaces}
\addtolength{\belowdisplayskip}{\extraspaces}
\addtolength{\abovedisplayshortskip}{\extraspace}
\addtolength{\belowdisplayshortskip}{\extraspace}}
\newcommand{\ee}{\end{equation}}
\newcommand{\ba}{\begin{eqnarray}
\addtolength{\abovedisplayskip}{\extraspaces}
\addtolength{\belowdisplayskip}{\extraspaces}
\addtolength{\abovedisplayshortskip}{\extraspace}
\addtolength{\belowdisplayshortskip}{\extraspace}}
\newcommand{\ea}{\end{eqnarray}}
\newcommand{\nonu}{\nonumber \\[.5mm]}
\newcommand{\A}{&\!\!\!}
\begin{document}
\thispagestyle{empty}
\setlength{\baselineskip}{6mm}
\begin{flushright}
SIT-LP-11/13\\
November, 2013
\end{flushright}
\vspace{7mm}
\vspace*{10mm}
\begin{center}
{\large \bf On chiral symmetry in extended nonlinear supersymmetry}
\\[15mm]
{\sc Kazunari Shima}
\footnote{
\tt e-mail: shima@sit.ac.jp} \ 
and \ 
{\sc Motomu Tsuda}
\footnote{
\tt e-mail: tsuda@sit.ac.jp} 
\\[2mm]
{\it Laboratory of Physics, 
Saitama Institute of Technology \\
Fukaya, Saitama 369-0293, Japan}
\vspace{1cm}

\begin{abstract}
We study the equations of the motion for Nambu-Goldstone (NG) fermions in the nonlinear representation of supersymmetry (NLSUSY) 
and show in $N$-extended NLSUSY that those equations under massless conditions are satisfied, provided the NG fermions are chiral eigenstates. 
The physical significance of the result is discussed in the context of NLSUSY general relativity 
through the relations between nonlinear and linear SUSY theories, 
which gives a new insight into the chiral symmetry for the low energy particle physics. 
\\[7mm]
%
%
\noindent
PACS: 11.30.Rd, 11.30.Pb, 12.60.Jv, 12.60.Rc \\[2mm]
\noindent
Keywords: chiral symmetry, supersymmetry, Nambu-Goldstone fermions, composite unified theory 
\end{abstract}
\end{center}

\newpage

\noindent
Supersymmetry (SUSY) is realized in linear (L) \cite{WB,WZ} and nonlinear (NL) \cite{VA} representations, 
which are related to each other by expressing component fields of LSUSY multiplet 
as SUSY invariant functionals of Nambu-Goldstone (NG) fermions of the Volkov-Akulov (VA) NLSUSY theory \cite{IK}-\cite{UZ}. 
The space-time structure of NLSUSY producing the massless NG fermions is generalized to curved space-time 
in nonlinear supersymmetric general relativity (NLSUSYGR) with a global SUSY algebra \cite{KS1,ST1}, 
whose tangent space-time is specified by means of $SL(2,{\bf C})$ Grassmann coordinates for the NG fermions 
corresponding to super $GL(4,{\bf R})$/$GL(4,{\bf R})$ in addition to $SO(3,1)$ Minkowski ones. 
In the generalized curved space-time,  the Einstein-Hilbert (EH) type action is obtained in terms of a unified vierbein 
which are composed of the vierbein of GR and the NG fermions. 
Corresponding to the spotaneous space-time (geometrical) SUSY breaking (called {\it Big Decay}), 
the NLSUSYGR action breaks into Riemann space-time, i.e. the EH action of GR from the scalar curvature term, 
the matter, i.e. the VA NLSUSY action from the cosmological term and higher-order interactions 
among the fundamental constituents of NLSUSYGR, i.e. those of the spin-2 graviton and the spin-1/2 NG fermions. 

The low energy physics induced from NLSUSYGR is realized 
by means of the relation between the NLSUSY theory and broken LSUSY ones (NL/LSUSY relation). 
Namely, NLSUSYGR is regarded as a fundamental theory behind (beyond) the standard model (SM) 
and it gives specific picture of basic fields in the true vacuum for the low energy (effective) theory. 
Each field of LSUSY multiplet is realized through the SUSY invariant relations as composites of the NG fermions 
(called {\it superons} tentatively in superon-graviton model (SGM) \cite{KS2}). 
Therefore, the solutions of equations of motion for the NG fermions in the NLSUSY(GR) theory 
would give important insights to the various basic results and assumptions for the low energy effective theory (and the SM) 
from the NLSUSYGR/SGM scenario. 

In our previous work \cite{STO} we have studied the equation of motion for the NG fermion in $N = 1$ VA NLSUSY theory 
and found that it is satisfied under a massless condition, provided the NG fermion is a {\it chiral} (left {\it or} right) eigenstate 
because of the higher order self-interaction terms with the NLSUSY structure. 
In this letter we extend the work to $N$-extended SUSY and study the equations of motion for the NG fermions in flat space-time. 
We show that they describe the massless chiral eigenstates for the NG fermions as in the case of $N = 1$ NLSUSY theory, 
which originate from the higher-order self-interaction terms, in particular, a third order interaction term of the NG fermions. 

Let us first show the VA NLSUSY action \cite{VA} for $N$-extended SUSY, 
\ba
L_{\rm NLSUSY} \A = \A - {1 \over {2 \kappa^2}} \vert w \vert 
\nonu
\A = \A - {1 \over {2 \kappa^2}} 
\left\{ 1 + t^a{}_a + {1 \over 2!} (t^a{}_a t^b{}_b - t^a{}_b t^b{}_a) 
\right. 
\nonu
\A \A 
\left. 
- {1 \over 3!} \epsilon_{abcd} \epsilon^{efgd} t^a{}_e t^b{}_f t^c{}_g 
- {1 \over 4!} \epsilon_{abcd} \epsilon^{efgd} t^a{}_e t^b{}_f t^c{}_g t^d{}_h 
\right\}, 
\label{NLSUSY}
\ea
where $\vert w \vert = \det w^a{}_b = \det(\delta^a_b + t^a{}_b)$, 
\be
t^a{}_b = - i \kappa^2 \bar\psi^i \gamma^a \partial_b \psi^i, 
\ee
with a dimensional constant $\kappa$ whose dimension is $(length)^2 = M^{-2}$ and $\psi^i$ represent $N$ NG Majorana fermions. 
\footnote{
Minkowski space-time indices are denoted by $a, b, \cdots = 0, 1, 2, 3$ 
and the metric is $\eta^{ab} = {\rm diag}(+,-,-,-)$. 
Latin indices $i, j, \cdots$ run from $1$ to $N$. 
Gamma matrices satisfy $\{ \gamma^a, \gamma^b \} = 2 \eta^{ab}$ 
and $\displaystyle{\sigma^{ab} = {i \over 4}[\gamma^a, \gamma^b]}$. 
}
The action (\ref{NLSUSY}) is invariant under NLSUSY transformations, 
\be
\delta_\zeta \psi^i = {1 \over \kappa} \zeta^i 
- i \kappa \bar\zeta^j \gamma^a \psi^j \partial_a \psi^i 
\ee
with constant (Majorana) spinor parameters $\zeta^i$, which 
satisfy a closed off-shell (SUSY) commutator algebra, 
$[\delta_{\zeta_1}, \delta_{\zeta_2}] = \delta_P(\Xi^a)$, 
where $\delta_P(\Xi^a)$ means a translation with parameters $\Xi^a = 2 i \bar\zeta_1^i \gamma^a \zeta_2^i$. 

The variation of the NLSUSY action (\ref{NLSUSY}) with respect to $\bar\psi^i$ gives 
the equations of motion for the NG fermions $\psi^i$ as 
\ba
\A \A 
\!\!\not\!\partial \psi^i 
+ \left\{ t^a{}_a \!\!\not\!\partial \psi^i - t^a{}_b \gamma^b \partial_a \psi^i 
+ {1 \over 2} (\partial_a t^b{}_b - \partial_b t^b{}_a) \gamma^a \psi^i \right\} 
\nonu
\A \A 
- {1 \over 2} \epsilon_{abcd} \epsilon^{efgd} 
(t^a{}_e t^b{}_f \gamma^c \partial_g \psi^i + t^a{}_e \partial_g t^b{}_f \gamma^c \psi^i) 
\nonu
\A \A 
- {1 \over 12} \epsilon_{abcd} \epsilon^{efgh} 
(2 t^a{}_e t^b{}_f t^c{}_d \gamma^d \partial_h \psi^i 
+ 3 t^a{}_e t^b{}_f \partial_h t^c{}_g \gamma^d \psi^i) 
= 0. 
\label{eom}
\end{eqnarray}
%
%
Here we notice in the equations of motion (\ref{eom}) that each term is factorized in terms of \ $\!\!\not\!\partial \psi^i$ 
and/or an ${\cal O}(\psi^3)$ term $t^a{}_b \gamma^b \partial_c \psi^i$. 
By using the Fiertz transformations and the Clifford algebras of $\gamma$ matrices 
we can show that the term $t^a{}_b \gamma^b \partial_c \psi^i$ is expressed by means of {\it only} 
\ba
\A \A 
\bullet \ {\rm the\ terms\ proportional\ to}\ \!\!\not\!\partial \psi^i, 
\label{type1}
\\
\A \A 
\bullet \ {\rm the\ terms\ with}\ J^P = 0^\pm, 2^\pm \ {\rm tensor\ structure}, 
\nonu
\A \A 
\ \ \ {\rm which\ vanish\ for\ chiral\ eigenstates}\ \psi^i_L \ or \ \psi^i_R. 
\label{type2}
\ea
%
%
Indeed, it is rewritten as 
\be
t^a{}_b \gamma^b \partial_c \psi^i 
= {i \over 4} \kappa^2 \bar\psi^j \gamma^a \gamma_A \partial_c \psi^i \ \gamma^b \gamma^A \partial_b \psi^j, 
\label{psi3}
\ee
where we take the summation of $\gamma_A = {\bf 1}, i \gamma_5, \gamma_d, \gamma_5 \gamma_d, \sigma_{de}$ 
and $\gamma^A = {\bf 1}, -i \gamma_5, \gamma^d$, $-\gamma_5 \gamma^d, 2 \sigma^{de}$, respectively. 
In Eq.(\ref{psi3}) the first four terms are just the avobementioned terms (\ref{type1}) and (\ref{type2}). 
On the other hand, when we express the last term 
\be
{i \over 2} \kappa^2 \bar\psi^j \gamma^a \sigma_{de} \partial_c \psi^i \ \gamma^b \sigma^{de} \partial_b \psi^j 
= {i \over 2} \kappa^2 \bar\psi^j \gamma^a \sigma_{de} \partial_c \psi^i (\sigma^{de} \!\!\not\!\partial \psi^j - 2i \gamma^d \partial^e \psi^j), 
\label{psi3-2}
\ee
the second term in Eq.(\ref{psi3-2}) becomes 
\ba
\A \A 
\kappa^2 \bar\psi^j \gamma^a \sigma_{de} \partial_c \psi^i \ \gamma^d \partial^e \psi^j 
\nonu
\A \A = \kappa^2 \bar\psi^j \gamma_b \partial_c \psi^i \ \sigma^{ab} \!\!\not\!\partial \psi^j 
\nonu
\A \A 
\hspace{5mm} 
- {1 \over 4} \kappa^2 \epsilon^{abde} (\partial_c \bar\psi^i \gamma_b \gamma_f \partial_e \psi^j \ \gamma_5 \gamma_d \gamma^f \psi^j 
- \partial_c \bar\psi^i \gamma_5 \gamma_b \gamma_f \partial_e \psi^j \ \gamma_d \gamma^f \psi^j) 
\ea
after some calculations, whose structure is just the terms (\ref{type1}) and (\ref{type2}). 
Therefore, we conclude the equations of motion (\ref{eom}) are expressed only the two types of terms (\ref{type1}) and (\ref{type2}). 

This fact leads to the massless chiral eigenstates for the NG fermions which are described by Eq.(\ref{eom}). 
If we impose the massless conditions for $\psi^i$, 
\be
\!\!\not\!\partial \psi^i = 0, 
\label{massless}
\ee
then the equations of motion (\ref{eom}) are written only the terms of the type (\ref{type2}). 
This means that Eq.(\ref{eom}) is satisfied under the conditions (\ref{massless}), 
provided the NG fermions are {\it chiral} eigenstates $\psi^i_L \ or\ \psi^i_R$. 

Namely, the equations of motion in the VA NLSUSY theory for $N$-extended SUSY 
chose one chiral eigenstate $\psi^i_L \ or \ \psi^i_R$ satisfying the condition (\ref{massless}), 
which does not mean $\psi^i_R = 0$ or $\psi^i_L = 0$, respectively. 
This conclusion is not obtained by using the two component spinor formalism throughout the arguments. 

We summarize our results and briefly discuss the physical significance in the NLSUSYGR/SGM scenario. 
In this letter we have studied in general the equations of motion for the NG fermions in the VA NLSUSY theory for $N$-extended SUSY 
and found that they are expressed by means of only two types of the terms (\ref{type1}) and (\ref{type2}). 
As a {\it nontrivial} consequence we have shown that the equations of motion (\ref{eom}) describe 
the massless chiral eigenstates $\psi^i_L \ or \ \psi^i_R$ satisfying the condition (\ref{massless}). 
This fact means that the higher order self-interaction terms of $\psi^i$, 
in particular, the ${\cal O}(\psi^3)$ term $t^a{}_b \gamma^b \partial_c \psi^i$ 
constrains the chirality of the massless eigenstates for $\psi^i$ to either {\it left or right} of two chiral eigenstates, 
although the lowest order kinetic term \ $\!\!\not\!\partial \psi^i$ does not constrain the chirality by itself. 

In NL/L SUSY relation the massless fermions $\lambda^i$ in LSUSY multiplet are expressed in terms of the NG fermions, 
e.g. they are systematically represented as $\lambda^i = \xi \psi^i \vert w \vert$ with a dimensionless parameter $\xi$ 
in the case of a general vector supermultiplet for $N = 2$ LSUSY (QED) theory \cite{ST2}. 
Therefore, if these $\lambda^i$ in the true vacuum of NLSUSY \cite{ST3} 
are regarded as the chiral fermions (quarks and/or leptons) in the SM (MSSM), 
they are {\it chiral} because the constituents $\psi^i$ are chiral. 
Namely, if various component fields in LSUSY multiplet are regarded as the fundamental fields in the SM, 
which are expressed as massless composites of the NG fermions through the NL/L SUSY relation, 
then the dynamics of the NG fermions in the VA NLSUSY theory and in NLSUSYGR 
may give a new insight into the chiral symmetry and the chiral eigenstate for all massless fermions in the SM.

\newpage

%
\newcommand{\NP}[1]{{\it Nucl.\ Phys.\ }{\bf #1}}
\newcommand{\PL}[1]{{\it Phys.\ Lett.\ }{\bf #1}}
\newcommand{\CMP}[1]{{\it Commun.\ Math.\ Phys.\ }{\bf #1}}
\newcommand{\MPL}[1]{{\it Mod.\ Phys.\ Lett.\ }{\bf #1}}
\newcommand{\IJMP}[1]{{\it Int.\ J. Mod.\ Phys.\ }{\bf #1}}
\newcommand{\PR}[1]{{\it Phys.\ Rev.\ }{\bf #1}}
\newcommand{\PRL}[1]{{\it Phys.\ Rev.\ Lett.\ }{\bf #1}}
\newcommand{\PTP}[1]{{\it Prog.\ Theor.\ Phys.\ }{\bf #1}}
\newcommand{\PTPS}[1]{{\it Prog.\ Theor.\ Phys.\ Suppl.\ }{\bf #1}}
\newcommand{\AP}[1]{{\it Ann.\ Phys.\ }{\bf #1}}

\end{document}